\documentclass{jfm}

\usepackage{graphicx}
\usepackage{amsmath}
\usepackage{mathrsfs}
\usepackage{newtxtext}
\usepackage{multirow}
\usepackage{newtxmath}
\usepackage{natbib}
\usepackage{dsfont}
\usepackage{hyperref}
\usepackage{float}
\usepackage{tabularx}
\hypersetup{
    colorlinks = true,
    urlcolor   = black,
    citecolor  = black,
}

\newcommand{\RomanNumeralCaps}[1]

\newcommand{\avg}[1]{\left\langle #1 \right\rangle}

\title{Scalar mixing in non-Markovian homogeneous isotropic synthetic turbulence}

\author{Pratyush S. Awasthi,
  Joaquim P. Jossy, Amitabh Bhattacharya
 \and Prateek Gupta,
 \corresp{\email{prgupta@iitd.ac.in}}}

\affiliation{Department of Applied Mechanics, Indian Institute of Technology Delhi, Hauz Khas, New Delhi, 110017, India.}
\newcommand{\boldface}[1]{\boldsymbol{#1}}  

\newcommand{\bff}{\boldface{f}}

\newcommand{\bfk}{\boldface{k}}

\newcommand{\bfr}{\boldface{r}}

\newcommand{\bfu}{\boldface{u}}

\newcommand{\bfx}{\boldface{x}}
\newcommand{\bfy}{\boldface{y}}

\newcommand{\bfF}{\boldface{F}}

\newcommand{\bfX}{\boldface{X}}
\newcommand{\bfY}{\boldface{Y}}

\newcommand{\bfalpha}{\boldsymbol{\alpha}}

\newcommand{\bfeta}{\boldsymbol{\eta}}

\begin{document}

\maketitle

\begin{abstract}
We show that non-Markovianity of the velocity field is an essential property of turbulent mixing. We demonstrate this via passive scalar mixing by synthetically generated stochastic velocity fields. Including a separate velocity decorrelation time scale for each spatial scale (random sweeping) yields an essentially non-Markovian velocity field with a finite time memory decaying as $\tau^{-5}$ (for a decaying spectrum) instead of an exponential decay (Markovian), which is obtained by including a constant time scale for all spatial scales, irrespective of the filtering function. We characterize the Lagrangian mixing statistics of both the Markovian and non-Markovian synthetic fields and compare them against a corresponding incompressible direct numerical simulation (DNS). We also study diffusive passive scalar mixing in  the Schmidt number range $\mathrm{Sc}\leq 1$ using the DNS and the synthetic fields. While both the synthetic fields recover the $-17/3$ scalar spectrum for low Schmidt numbers, the mean gradients in a decaying simulation, as well as the production and dissipation of scalar variance in a statistically stationary simulation, are severely underpredicted by the Markovian fields compared to the non-Markovian fields. Throughout, we compare our results with companion 3D DNS to show the necessity of non-Markovianity in synthetic fields to capture mixing dynamics.
\end{abstract}
 

\section{Introduction}\label{sec:Introduction}

Mixing refers to the dynamic process by which the scalar gradients are redistributed and homogenized through the combined action of stirring and molecular diffusion \citep{zhou2024hydrodynamic,jossy2025mixing}. Action of stirring by turbulence can be studied in statistically stationary systems with imposed concentration gradients~\citep{yeung2014direct}, as well as decaying systems, in which the homogenization of an initially inhomogeneous field of concentration is studied~\citep{yeung2013spectrum}. In steady mixing setups, the balance between stirring and diffusion signifies mixing. In contrast, unsteady mixing setups are characterised by stirring dominated mixing followed by diffusion dominated smoothening. In this study, we elucidate the effect of temporal correlations on the mixing dynamics of both steady and unsteady mixing setups.

Advection of small-scale eddies in a turbulent flow by large, energy-containing eddies is called the random sweeping effect~\citep{tennekes1975eulerian}. Using DNS, \cite{gorbunova2021spatio} showed that the decorrelation time decays as $k^{-1}$ at large wavenumbers, rather than the classical $k^{-2/3}$, indicating that the large-scale eddies determine the temporal decorrelation.
For homogeneous isotropic turbulence (HIT),~\cite{tennekes1975eulerian}  demonstrated that the Eulerian time spectrum exceeds its Lagrangian counterpart at high frequencies. Building on this idea, \cite{yeung2002random} used the random-sweeping framework to elucidate Lagrangian statistics relevant to mixing and dispersion, and showed that the concept extends to passive scalars. A Lagrangian measure of enhanced stirring is the exponential stretching captured by local finite-time Lyapunov exponents (FTLEs) \citep{aref2020stirring,toussaint2000spectral}. \cite{gotzfried2019comparison} compared passive-scalar mixing in both Eulerian and Lagrangian descriptions and showed that the mean mixing time -- defined as the duration over which stirring dominates in unsteady mixing setups -- is governed by the compressive local FTLE. 

Synthetic turbulent fields aim to reproduce the essential dynamical features of turbulence while avoiding the computational cost associated with solving the full Navier–Stokes equations. Beyond practical utility, such models provide valuable insights into the intrinsic mechanisms of turbulence \citep{juneja1994synthetic}.  \cite{batchelor1959small2} predicted that, at high Reynolds numbers and low scalar diffusivity, HIT produces a passive-scalar spectrum with the characteristic $-17/3$ scaling in the diffusive (viscous–convective) range.  \cite{holzer1994turbulent} demonstrated that two-dimensional, stochastically generated Gaussian velocity fields are able to reproduce this $-17/3$ scaling when used to mix passive scalars.
\cite{jossy2025active} extended this synthetic-field framework to three dimensions, but found that such fields do not replicate the stirring efficacy of Navier-Stokes-generated turbulent flows. In this work, we formulate the synthetic generation of homogeneous isotropic turbulent fields capable of replicating the mixing efficiency of non-synthetic turbulent fields.
In this work, we show that non-Markovian three-dimensional homogeneous synthetic velocity fields are closer to turbulence in mixing passive scalars than earlier developed Markovian fields. In the next section, we outline the theoretical basis for constructing non-Markovian synthetic fields
 and describe the numerical implementation. We then discuss the results in \S~\ref{sec: results} followed by a summary of findings in \S~\ref{sec: Conclusions}.
\section{Theory and numerical simulations}\label{numerics}

\subsection{3D time correlated non-Markovian synthetic field}\label{sec: 3D_synthetic_field}
\cite{careta1994diffusion} proposed a two dimensional homogeneous isotropic velocity field using the Ornstein-Uhlenbeck process. \cite{jossy2025active} extended the methodology to generate three dimensional homogeneous isotropic velocity field. In this work, we extend the 3D synthetic velocity field proposed by \cite{jossy2025active} to generate a 3D non-Markovian homogeneous isotropic velocity field. We define a vector potential $\hat{\boldsymbol{{\eta}}}(\boldsymbol{k},t)$ in the spectral space which is an Uhlenbeck-Ornstein process with different decaying time scales given as,
\begin{equation}
    \frac{d\hat{\boldsymbol{\eta}}(\boldsymbol{k},t)}{dt} = - \frac{\hat{\boldsymbol{\eta}}(\boldsymbol{k},t)}{\tau(\boldsymbol{k})} + \frac{\hat{Q}(\boldsymbol{k})}{\tau(\boldsymbol{k})}\hat{\boldsymbol{\chi}}(\boldsymbol{k},t),
    \label{eq: OU_spectral}
\end{equation}
where $\tau(\boldsymbol{k})$ is the spectrally varying time scale, $\hat{Q}(\boldsymbol{k})$ is the filtering kernel used to achieve a desired energy spectrum, and $\hat{\boldsymbol{\chi}}(\boldsymbol{k},t)$ is the Fourier transform of $\boldsymbol{\chi}(\boldsymbol{r},t)$ which is a delta correlated white noise vector with {zero mean} and its correlation tensor is defined as
\begin{equation}
    \langle \chi_i(\boldsymbol{x},t) \chi_j(\boldsymbol{x}+\bfr,t+s) \rangle = 2\epsilon \delta_{ij} \delta(\bfr)\delta(s), 
\end{equation}
where $\epsilon$ is the intensity of the white noise and $\hat{()}$ denotes the respective quantities in the Fourier space. The solution of the vector potential in the Fourier space is given as
\begin{equation}
    \hat{\boldsymbol{\eta}}(\boldsymbol{k},t) = \hat{\boldsymbol{\eta}}(\boldsymbol{k},0)e^{-\frac{t}{\tau(\boldsymbol{k})}} + \int_0^t \frac{{\hat{Q}}(\boldsymbol{k})}{\tau(\boldsymbol{k})}\hat{\boldsymbol{\chi}}(\boldsymbol{k},t') e^{\frac{t-t'}{\tau(\boldsymbol{k})}} dt',
    \label{eq: OU_spectral_solution}
\end{equation}
with the time correlation tensor defined as,
\begin{align}
\langle \hat{\eta}^*_p(\boldsymbol{k},t)\hat{\eta}_q(\boldsymbol{k},t+s) \rangle  &= |\hat{Q}|^2 \frac{2\epsilon \delta_{pq} e^{-s/\tau(\boldsymbol{k})}}{(2\pi)^3 \tau(\boldsymbol{k})^2} \int_0^t \int_0^{t+s} \delta(t'-t'') e^{\frac{t'+t'' - 2t}{\tau(\boldsymbol{k})}} \, dt' \, dt''.
\end{align}
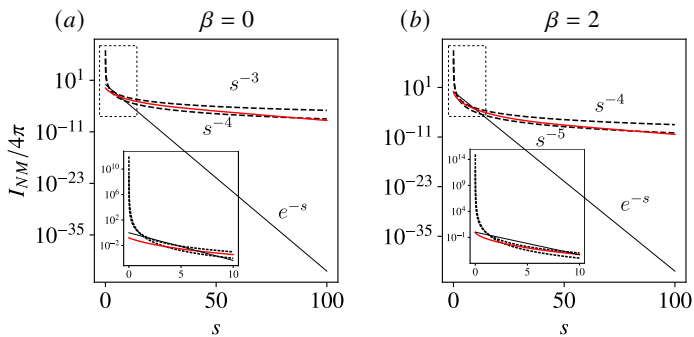
\begin{figure}
\centering
\includegraphics[width=0.67\textwidth]{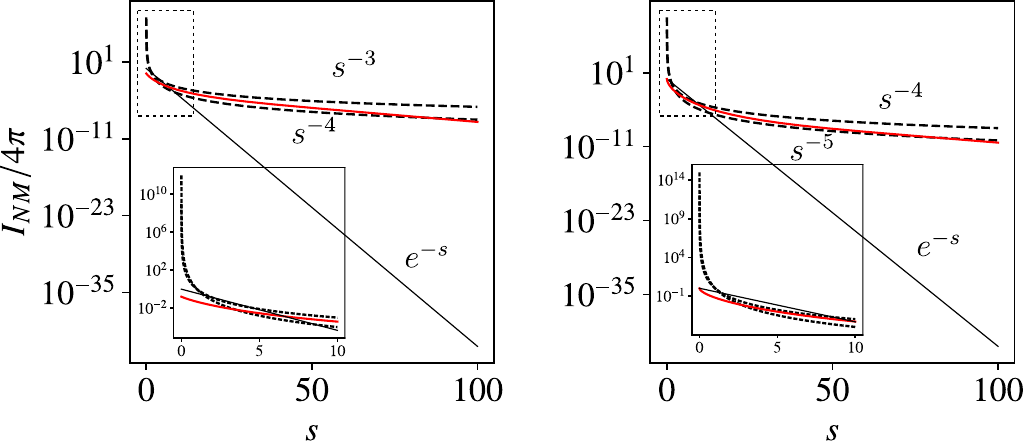}
\put(-240,115){$(a)$}
\put(-185,115){$\beta=0$}
\put(-110,115){$(b)$}
\put(-55,115){$\beta=2$}
\caption{Comparison of the time correlation integral in \eqref{eq: time-correlation-integral} with $e^{-s}$, $1/s^3$, and $1/s^4$. Exponentially decaying time correlation corresponds to Markovian vector potential. Clearly, \eqref{eq: OU_spectral} defines a non-Markovian vector potential.}
\label{fig: time_correlation_NM}
\end{figure}
When, $t\to\infty$, the above expression yields, 
\begin{equation}
\lim_{t\to\infty}\langle{\hat{\eta}^*_p(\boldsymbol{k},t)\hat{\eta}_q(\boldsymbol{k},t+s)}\rangle = \frac{2\epsilon\delta_{pq}|\hat{Q}|^2}{(2\pi)^3\tau(\boldsymbol{k})}e^{-s/\tau(\boldsymbol{k})}.
\label{eq: eta_correlation_result}
\end{equation}
To estimate the velocity auto-correlation time scale (decorrelation time for a particular scale) for synthetic fields, we substitute the above expression in
\begin{equation}
     \phi_{ii}(\boldsymbol{k},s) = \left\langle\hat{\boldsymbol{u}}^*(\boldsymbol{k},t)\cdot\hat{\boldsymbol{u}}^*(\boldsymbol{k},t+s)\right\rangle = k^2\left\langle\hat{\eta}^*_p\hat{\eta}_q\right\rangle - k_mk_l\left\langle\hat{\eta}^*_m\hat{\eta}_l\right\rangle,
\end{equation}
to obtain
\begin{equation}
    \tau_c = \frac{\int^\infty_0 \phi_{ii}(\boldsymbol{k},s)ds}{\phi_{ii}(\boldsymbol{k},0)} = \tau(\boldsymbol{k}).
    \label{eq: decorrelation}
\end{equation}
Equation~\eqref{eq: decorrelation} confirms that the decorrelation time scale of the velocity field at length scale $1/|\bfk|$ is quantified by $\tau(\bfk)$. The auto-correlation of the Fourier transform of velocity $\phi_{ii}(\boldsymbol{k},s)$ is related to the energy spectra $E(k)$ as~\citep{batchelor1953theory}, 
\begin{equation}
     E(k) = 2\pi k^2 \phi_{ii} = \frac{k^4 \epsilon|\hat{Q}|^2}{\pi^2\tau},
     \label{eq: correlation-spectrum-relation}
\end{equation}
where $k=|\boldsymbol{k}|$. A desired spatial variation in the kinetic energy spectrum is obtained using the filtering kernel $\hat{Q}(\boldsymbol{k})$ defined as
\begin{equation}
    \hat{Q}_I(\boldsymbol{k}) = (1+\ell^2 k^2)^{-n},
    \label{eq: Q_operator}
\end{equation}
where $\ell$ denotes a length scale. The subscript $()_I$ denotes that this form of $\hat{Q}$ is used for achieving the so-called ideal spectra in our synthetic fields, i.e. $E(k)\sim k^{-5/3}$ for $\ell k \gg 1$. For varying $\tau(\boldsymbol{k})$ (not constant), the process $\bfeta(\boldsymbol{x},t)$ is a non-Markovian process. In the spatial domain, the time correlation is given by, 
\begin{equation}
    \left\langle \eta_p(\boldsymbol{x},t)\eta_q(\boldsymbol{x}+\bfr,t+s)\right\rangle = \frac{1}{(2\pi)^3}\int_{\mathbb{R}^3}\left\langle\hat{\eta}^*_p(\boldsymbol{k},t)\hat{\eta}_q(\boldsymbol{k},t+s)\right\rangle e^{i\bfk\cdot\bfr}d\boldsymbol{k},
\end{equation}
which yields for $\hat{Q}_I(\boldsymbol{k})$ from \eqref{eq: Q_operator},
\begin{equation}
    C^\eta_{pq}(\bfr, s) = \left\langle\eta_p(\bfx,t)\eta_q(\bfx+\bfr,t+s)\right\rangle = \frac{2\epsilon \delta_{pq}}{(2\pi)^3}\int_{\mathbb{R}^3}\frac{(1+\lambda^2 k^2)^{-2n}}{\tau(\bfk)}e^{-s/\tau(\bfk)}e^{i\bfk\cdot\bfr}d\bfk,
    \label{eq: time_corr_real_space}
\end{equation}
which resembles the correlation field studied by~\cite{chaves2003lagrangian}. 
\begin{figure}
\centering
\includegraphics[width=\textwidth]{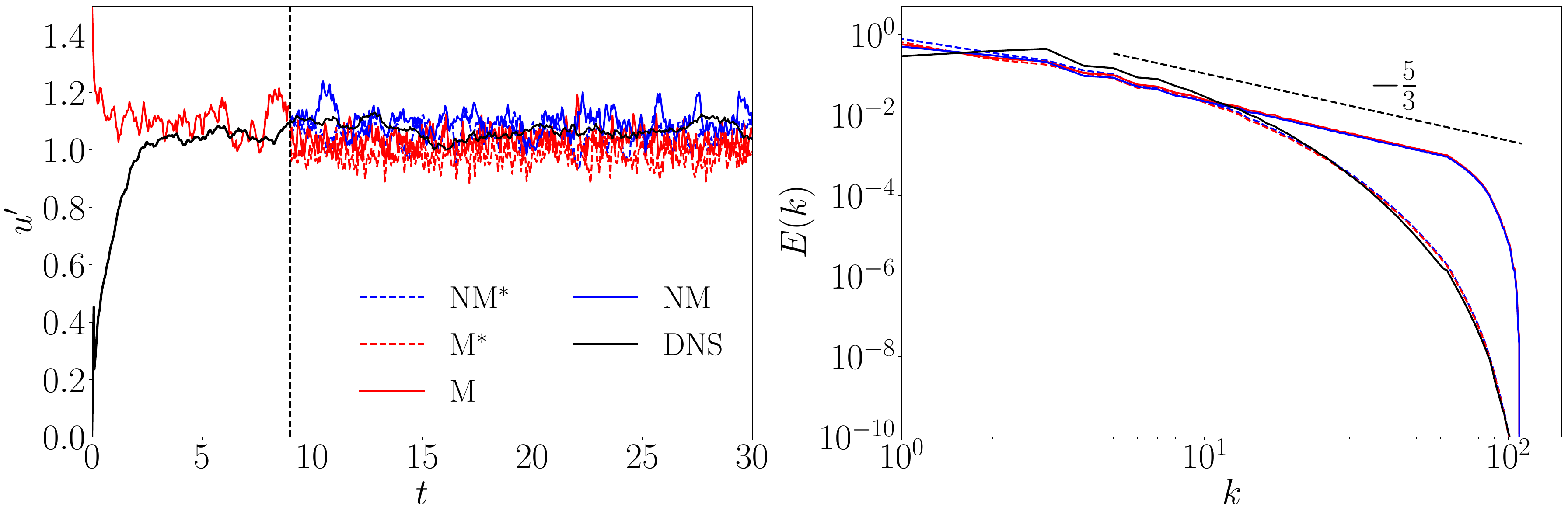}
\put(-385,125){$(a)$}
\put(-190,125){$(b)$}
\caption{Comparison of (a) root mean square evolution for different velocity fields. (b) kinetic energy spectra for different velocity fields. For DNS, $k_\mathrm{max}\eta \approx 1.76$. The legends are the same for both figures. }
\label{fig: urms_spectra}
\end{figure}

\begin{table}
\centering
\begin{tabular}{lccc}
 & \textbf{DNS} & \textbf{Markovian} & \textbf{Non-Markovian} \\
\textbf{Steady}     & DS1, DS2  & MS1$^{*}$, MS2, MS4, MS8, MS16$^{*}$ & NMS1$^{*}$, NMS2, NMS4, NMS8, NMS16$^{*}$ \\
\textbf{Decaying}   & DU1, DU2  & MU1$^{*}$, MU2, MU4, MU8, MU16$^{*}$ & NU1$^{*}$, NMU2, NMU4, NMU8, NMU16$^{*}$ \\
\end{tabular}
\caption{Parameter space for the simulations for both DNS and synthetic turbulent fields. The $()^{*}$ indicates that the simulations were run for both ideal and matched spectra. \textcolor{black}{The numbers in case names are equal to $1/\mathrm{Sc}$ of the scalar.}}
\label{tab:Synthetic-cases}
\end{table}
Equation~\eqref{eq: time_corr_real_space} highlights that for varying $\tau(\boldsymbol{k})$, at a spatial point, $\boldsymbol{\eta}$ is a non-Markovian field since the integral can not be simplified to an exponentially decaying time correlation, unless $\tau(\boldsymbol{k})=\mathrm{constant}$ (the Markovian case). As we show in~\S~\ref{sec: results}, the velocity field generated due to Markovian $\boldsymbol{\eta}$ leads to weak scalar mixing, while the non-Markovian velocity field exhibits mixing dynamics very similar to turbulence. As shown in \eqref{eq: decorrelation}, $\tau(\boldsymbol{k})$ is the time after which velocity at scale $1/|\boldsymbol{k}|$ gets de-correlated with its history. In the context of hydrodynamic turbulence, $\tau(\bfk)$ can be compared to the residence time of velocity perturbations at a particular length scale, before they get completely renewed. While the time scale for eddies at wavenumber $k$ can be shown to be $\sim k^{-\frac{2}{3}}$~\citep{pope2001turbulent}, the so-called random sweeping effect \citep{gorbunova2021spatio, yeung2002random,wilczek2012wave} results in $\tau\sim \left(u_0 k\right)^{-1}$ where $u_0$ is the large eddy velocity scale. Using a radial function for $\tau(\bfk) = 1/k$, and $n=5/3$ in \eqref{eq: Q_operator}, we can simplify the integral in time correlations of $\bfeta, \bfu, $and $\nabla\bfu$ as, 
\begin{equation}
    I_{NM}(s) = \frac{4\pi}{C}\int^\infty_0k^{\beta+3}\left(1 + \lambda^2k^2\right)^{-\frac{10}{3}}e^{-ks}dk,
    \label{eq: time-correlation-integral}
\end{equation}
where $\beta = 0, 2, 4$ for $\eta, \bfu$, and $\nabla \bfu$, respectively. Figure~\ref{fig: time_correlation_NM} shows the variation of $I_{NM}/4\pi$ with $s$. The time correlation clearly decays slower than $e^{-s}$, thus confirming the non-Markovian nature of the $\bfeta$ field, and consequently, that of $\bfu$ and $\nabla \bfu$. 

\subsection{Numerical implementation and mixing simulations}
\label{sec: numerics}
Using the non-Markovian field developed above, and 3D incompressible DNS, we perform decaying and statistically stationary simulations of a passive scalar mixing. For numerical implementation of the synthetic fields, we follow~\cite{careta1994diffusion} and work with the time-discretized version of \eqref{eq: OU_spectral_solution} to obtain
\begin{equation}
    \hat{\bfeta}(\bfk,t+\Delta t) = \hat{\bfeta}(\bfk,t)e^{-\frac{\Delta t}{\tau(\bfk)}} + \hat{Q}(\bfk)\bfalpha(\bfk,t)\sqrt{\frac{\epsilon}{2\tau(\bfk)(2\pi)^3}\left(1 - e^{-\frac{2\Delta t}{\tau(\bfk)}}\right)},
\end{equation}
where $\bfalpha(\bfk)$ is also a delta-correlated Gaussian vector with unit intensity and $\hat{\bfeta}(\bfk,0) = \sqrt{\epsilon/\tau}\hat{Q}\hat{\chi}$. We perform two sets of synthetic velocity mixing simulations: Markovian synthetic velocity field ($\tau(\bfk)=0.1~(\mathrm{const.})$) and non-Markovian synthetic velocity field ($\tau(\bfk) = 1/k$). \textcolor{black}{The value $\tau=0.1$ for Markovian field is chosen to keep $E(k=1)$ similar to DNS values (c.f. figure~\ref{fig: urms_spectra}). \cite{jossy2025active} have shown that with $\tau = 0.01$ ($\hat{Q}$ can be adjusted to keep suitable values of $E(k)$ as per \eqref{eq: correlation-spectrum-relation}) fields mix very less. In the limit $\tau\to\infty$, we will obtain a steady field. It is noteworthy that the random sweeping approximation eliminates the arbitrariness related to the value of $\tau$.} For each of these sets, we run two further sets of simulations: statistically stationary scalar mixing with an imposed mean scalar gradient and decaying scalar mixing with an initialized spherical blob. To isolate the effect on mixing (if any), we further perform two sets of simulations for each: synthetic fields with ideal kinetic energy spectra of $k^{-5/3}$ using $\hat{Q}_I$ defined in \eqref{eq: Q_operator} and synthetic fields with kinetic energy spectra matched with our DNS simulations for which we use
\begin{equation}
    \hat{Q}_M = \hat{Q}_Ie^{-6.2\left(\frac{k-10}{64\sqrt{3} - 10}\right)^{1.05}}.
\end{equation}
\begin{figure}
\centering
\includegraphics[width=1.0\textwidth]{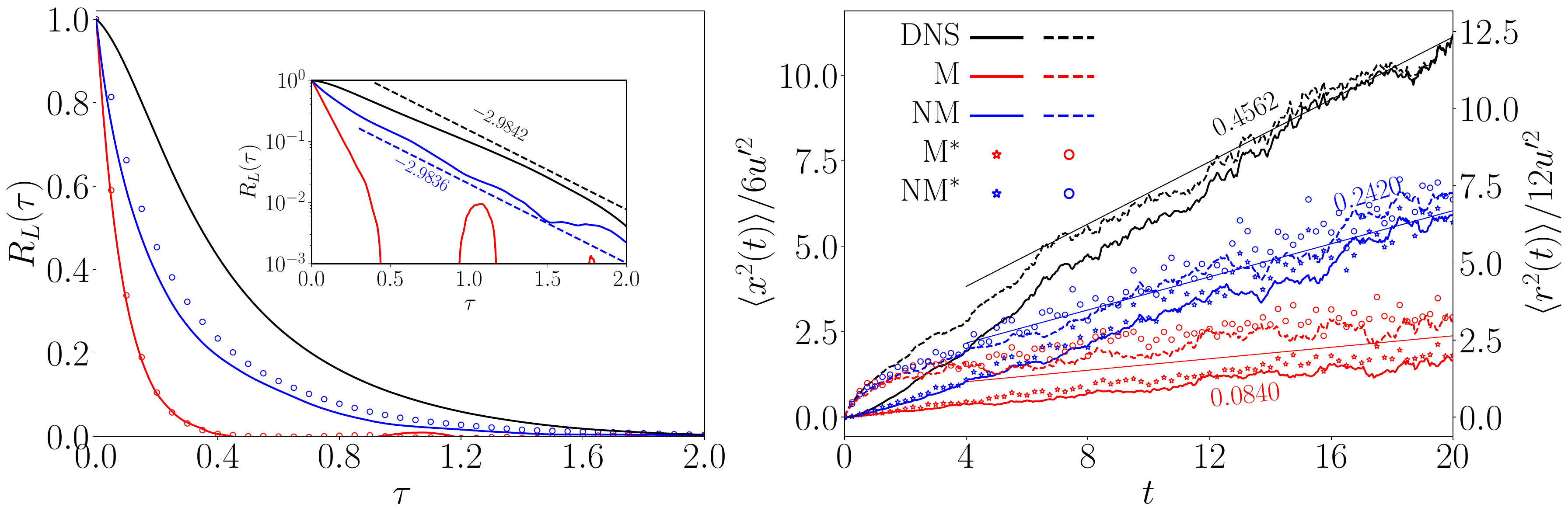}
\put(-380,130){$(a)$}
\put(-190,130){$(b)$}
\caption{$(a)$ The Lagrangian velocity correlations and $(b)$ single particle (solid lines, empty stars) and pair (dashed lines, empty circles) dispersions for the Markovian, the non-Markovian, and the incompressible forced HIT simulations. $T_L$ for DNS: 0.46, NM: 0.25, M: 0.08.}
\label{fig: lagrangian-stats}
\end{figure}
For comparing the mechanics of mixing by synthetic velocity fields, we perform 3D incompressible DNS of passive scalar mixing for both the decaying and statistically stationary mixing cases. We solve the forced incompressible Navier-Stokes equations,
\begin{subequations}
\begin{align}
    &\nabla\cdot\bfu = 0,~ \frac{D\bfu}{Dt} = -\nabla p + \frac{1}{\mathrm{Re}}\nabla^2\bfu + \bff.
    \label{eq: NS-Scalar}
\end{align}
\end{subequations}All the simulations presented are performed with a 3D parallelized Fourier pseudo-spectral solver using P3DFFT++ library~\citep{pekurovsky2012p3dfft} on a $192^3$ grid with 2/3 dealiasing ($k_{\mathrm{max}}=64$) in a $(2\pi)^3$ domain. In \S~\ref{sec: results}, we also discuss Lagrangian tracking statistics, for which we use Lagrangian particle tracking implemented using PETSc~\citep{may2017dmswarm}. The forcing $\bff$ exists only in small wavenumbers $1<k<2\sqrt{3}$, governed by a process similar to \eqref{eq: OU_spectral}~\citep{eswaran1988examination}. Synthetic velocity field simulations are done such that the turbulent kinetic energy $\kappa$ and $u'=\sqrt{2{\avg{\kappa}_x}/3}$ are similar as shown in figure~\ref{fig: urms_spectra}(a), where $\avg{}_x$ denotes spatial average. figure~\ref{fig: urms_spectra}(b) shows that the synthetic field is able to reproduce both the ideal and DNS energy spectra accurately.
Velocity fields obtained either from DNS or from synthetic fields are used for mixing scalars governed by the equation,
\begin{equation}
    \frac{D\phi}{Dt} = \frac{1}{\mathrm{Re}\mathrm{Sc}}\nabla^2\phi,
    \label{eq: scalar}
\end{equation}
where $\mathrm{Re}$ is a characteristic Reynolds number and $\mathrm{Sc}$ is the Schmidt number. We perform DNS for only $\mathrm{Sc}=1$ and $1/2$ to save computational cost. In both synthetic and DNS runs, we use $\mathrm{Re}=125$. 

\begin{figure}
\centering
\includegraphics[width=\textwidth]{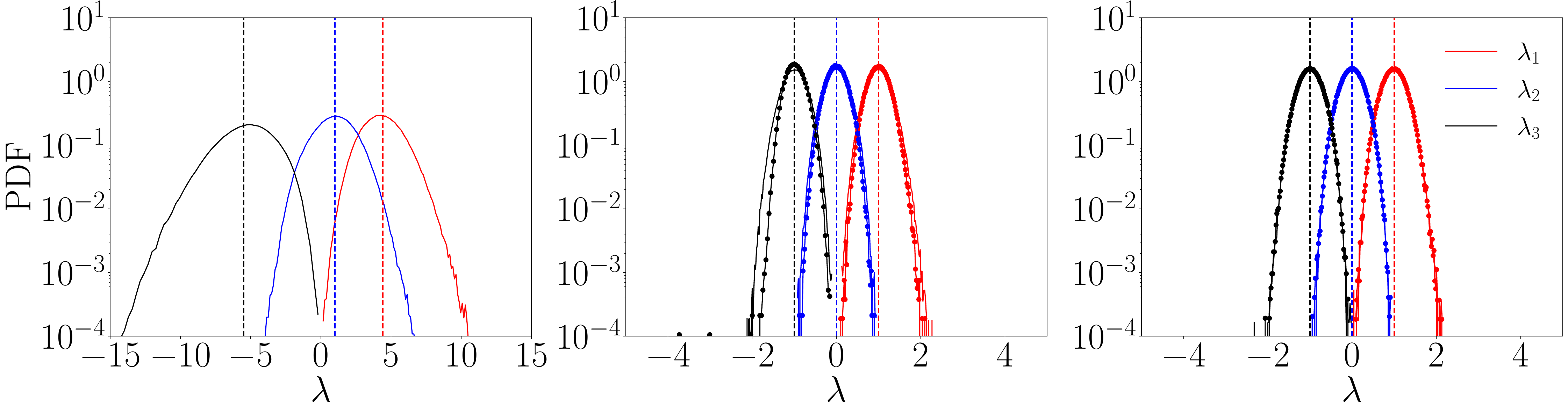}
\put(-380,105){$(a)$}
\put(-250,105){$(b)$}
\put(-125,105){$(c)$}
\put(-315,100){DNS}
\put(-185,100){NM, $\mathrm{NM}^*$}
\put(-62,100){NM, $\mathrm{M}^*$}
\caption{The distribution of the Lyapunov exponents for the DNS and the synthetic fields.}
\label{fig: lyapunov_exponent}
\end{figure}
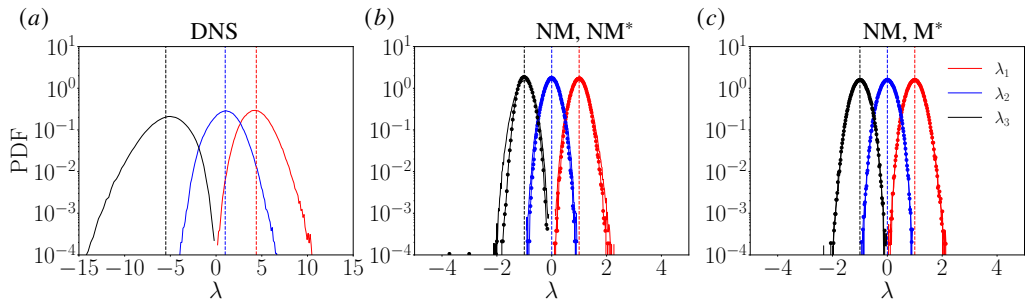
\section{Results}
\label{sec: results}
\subsection{Lagrangian statistics}
We calculate single particle correlation, pair dispersion, and the finite time Lyapunov exponents of Lagrangian particles in both the synthetic fields and compare them with those from DNS. Positions of the particles which follow the fluid streamlines are governed by,
\begin{equation}
    \bfx^+(t) = \bfX + \int^t_0\bfu(\bfx^+(s);s)ds, 
\end{equation}
where $\bfX$ denotes the initial location at $t=0$. The single particle and velocity correlations are given by~\citep{sawford20124},
\begin{align}
    \avg{\bfu^+(t)\cdot\bfu^+(t+\tau)} &= 2u'^2R_L(\tau),\\ \avg{\bfx^+(t)\cdot\bfx^+(t)} - \avg{\bfx^+(0)\cdot\bfx^+(0)}&= 6u'^2\int^t_0\int^{t'}_0 R_L(\tau) d\tau dt' = \avg{x^2(t)},
\end{align}
using which, the Lagrangian integral time scale is defined as,
\begin{equation}
    T_L = \int^\infty_0 R_L(\tau)d\tau.
\end{equation}
For $t\gg T_L$, the single particle correlation tends to $6u'^2T_Lt$. Similarly, pair dispersion can be computed using the separation between two particles, 
\begin{equation}
    \bfr^+ = \bfx^+-\bfy^+ = \bfX-\bfY + \int^t_0\left(\bfu(\bfx^+(s);s)-\bfu(\bfy^+(s);s)\right)ds.
\end{equation}
For $t\gg T_L$, the position of the two particles become independent, hence $\avg{|\bfr^+(t) - \bfr^+(0)|^2}\to 12u'^2 T_L = \avg{r^2(t)}$. Figure~\ref{fig: lagrangian-stats} summarizes these results. The non-Markovian velocity fields clearly exhibit Lagrangian integral time scales closer to the DNS compared to the Markovian fields for both ideal and matched spectra.

\begin{figure}
\centering
\includegraphics[width=\textwidth]{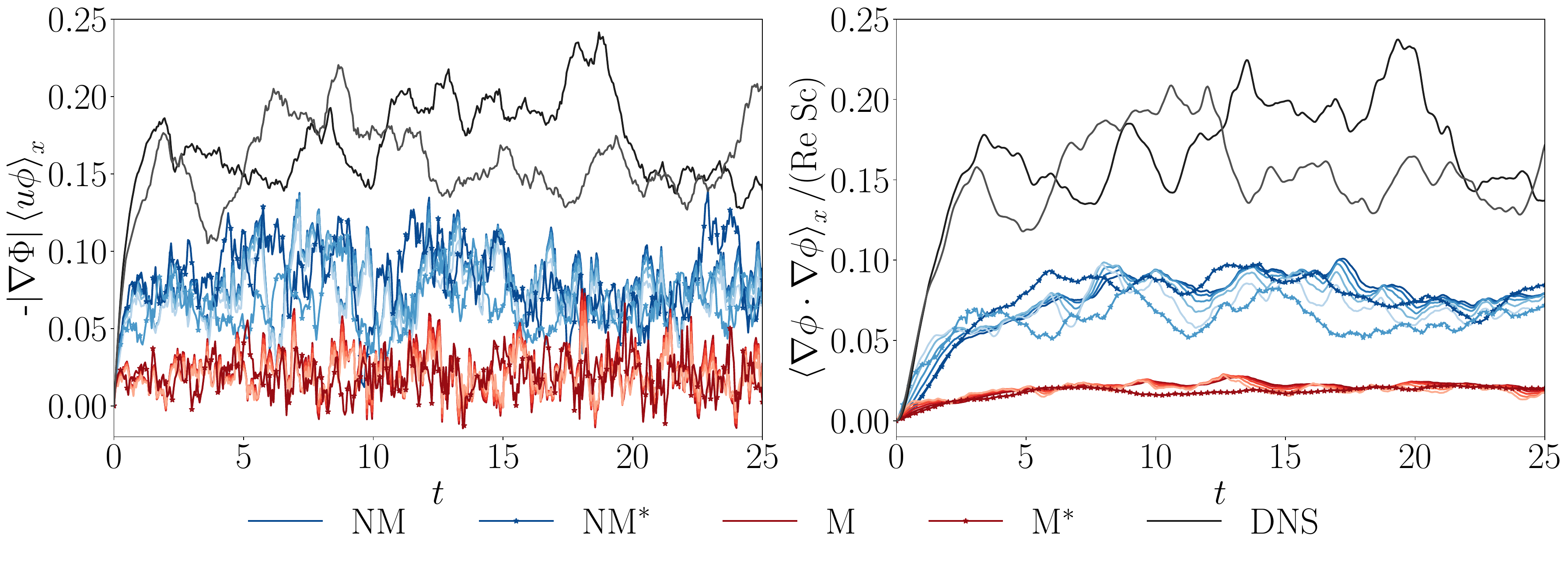}
\put(-380,145){$(a)$}
\put(-190,145){$(b)$}
\caption{Timeseries of spatial average of $(a)$ scalar production and $(b)$ scalar dissipation for steady scalar mixing using DNS, non-Markovian, and Markovian velocity field for all Schmidt number cases. }
\label{fig: mean_grad_dissipation_production}
\end{figure}
Based on the Lagrangian positions, we can define the deformation gradient at each particle~\citep{ottino1989kinematics} as,
\begin{equation}
    \bfF\left(\bfX, t\right) = \nabla_X\bfx^+,
\end{equation}
which is related to the velocity gradient as, 
\begin{equation}
    \frac{d\bfF}{dt} = \nabla\bfu^+ \cdot \bfF,
\end{equation}
where the Lagrangian velocity gradient is evaluated as $\nabla\bfu^+(t) = \nabla\bfu(\bfx^+(t);t)$. We track $\bfF(t)$ and its QR decomposition using a predictor-corrector time stepping~\citep{gotzfried2019comparison} and compute the Finite Time Lyapunov Exponents (FTLE) of the fields. Figure~\ref{fig: lyapunov_exponent} shows the distribution of the three Lyapunov exponents for the DNS and the synthetic fields. For DNS, the Lyapunov exponents approach the ratios $\avg{\lambda_1}:\avg{\lambda_2}:\avg{\lambda_3} \approx 4.3:1:-5.3$ using $64^3$ particles for averaging. As discussed by \cite{gotzfried2019comparison}, an HIT field tends to compress a scalar inhomongeneity in one direction, pulling it in other direction. The largest and smallest exponents correspond to these deformations. The middle exponent, which is proportional to the triple velocity correlations~\citep{balkovsky1999universal} is non-zero due to the intermittency in the DNS. The overall kinematics result in a sheet-like structure forming from an otherwise concentrated inhomogeneity. Even though, the dispersion characteristics of the non-Markovian synthetic fields are closer to the DNS than the Markovian fields, due to their inherent Gaussian nature, the Lyapunov exponents for both the synthetic fields approach the ratios $\avg{\lambda_1}:\avg{\lambda_2}:\avg{\lambda_3} \approx 1:0:-1$, irrespective of the spectra.
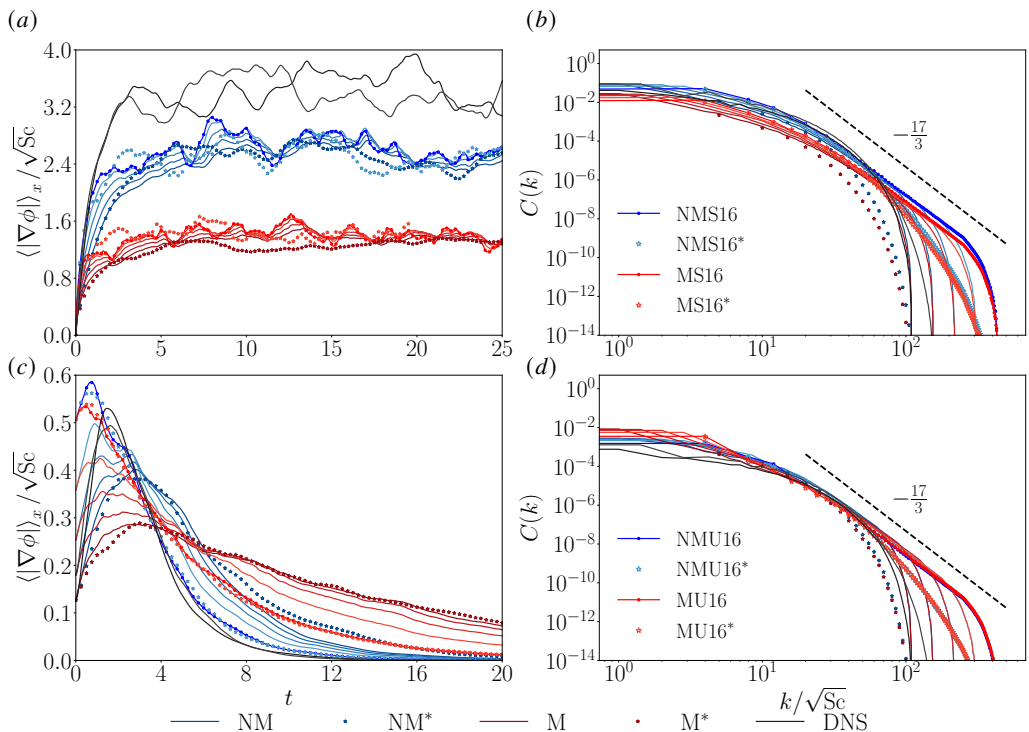
\begin{figure}
\centering
\includegraphics[width=\textwidth]{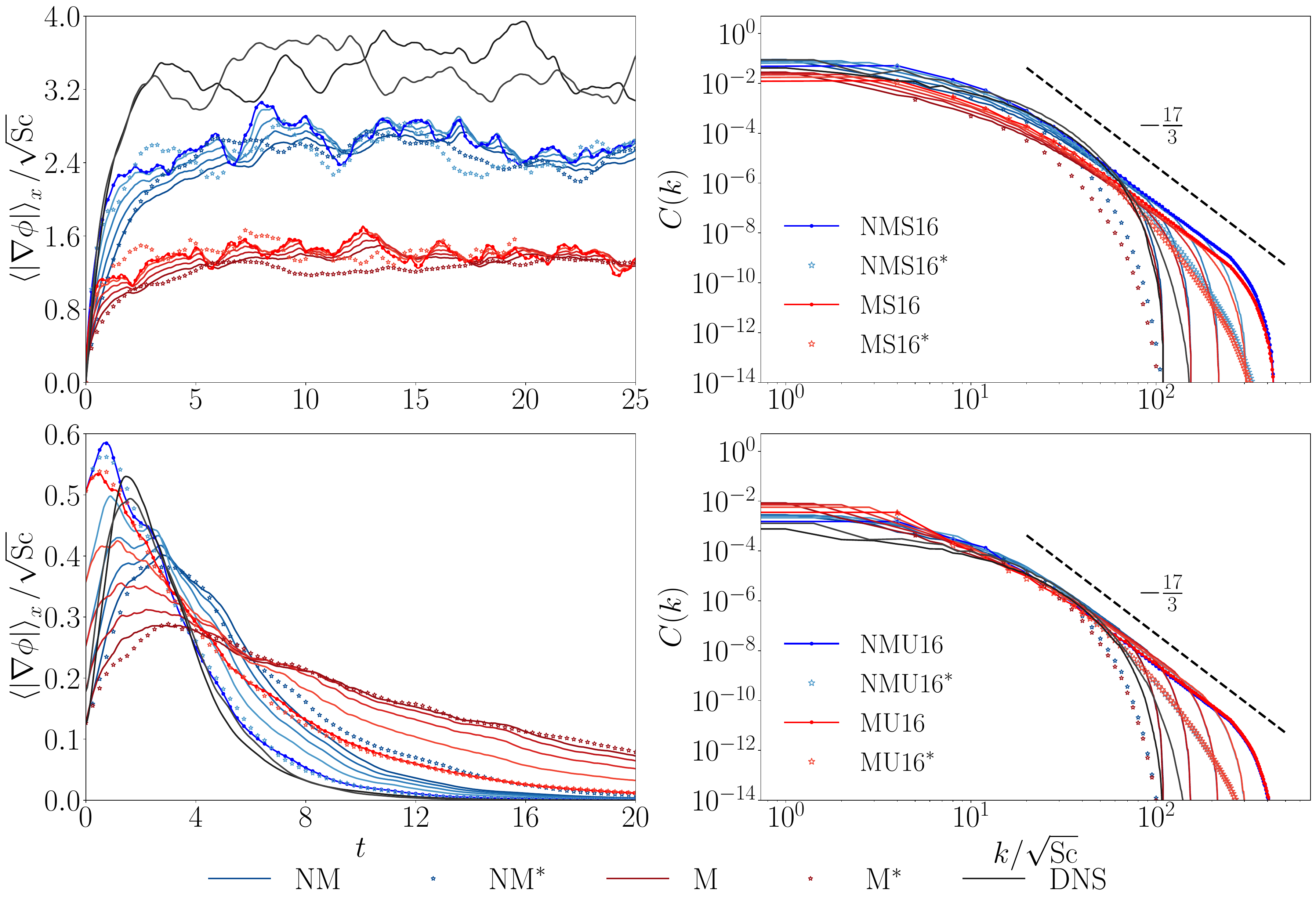}
\put(-385,270){$(a)$}
\put(-190,270){$(b)$}
\put(-385,140){$(c)$}
\put(-190,140){$(d)$}
\caption{Time series of $\avg{|\nabla \phi|}_x$ for $(a)$ stationary and $(c)$ decaying case; $(b)$ time-averaged scalar spectrum for steady scalar mixing and $(d)$ scalar spectrum at $t=3$ for decaying mixing with an arbitrary line of slope $-17/3$ for DNS, non-Markovian, Markovian velocity field for all Schmidt number cases.}
\label{fig: mean_grad_spectra_gradyc}
\end{figure}

\subsection{Statistically stationary scalar field}
We use the DNS and the synthetic fields to study mixing characteristics in a statistically stationary setup by solving for the concentration fluctuations assuming a constant mean gradient as,
\begin{equation}
    \frac{D\phi}{Dt} = -\bfu\cdot\nabla\Phi + \frac{1}{\mathrm{Re}\mathrm{Sc}}\nabla^2\phi.
    \label{eq: SteadyMeanGrad}
\end{equation}
The corresponding scalar fluctuation variance equation is~\citep{yeung2014direct}
\begin{equation}
 \frac{d }{d t}\frac{\avg{\phi^2}_x}{2} = -\avg{\bfu\phi}_x\cdot\nabla\Phi - \frac{\avg{|\nabla \phi|^2}_x}{\mathrm{Re}\mathrm{Sc}},
 \label{eq: production_dissipation}
\end{equation}
where $-\avg{\bfu\phi}_x\cdot\nabla\Phi$ is the production and ${\avg{|\nabla \phi|^2}_x}/{\mathrm{Re}\mathrm{Sc}}$ is the dissipation. In this work, we use $\nabla \Phi = (0.5,0,0)$. For spanning over different Schmidt numbers, we use the same synthetic field to save resources. The two DNS (DS1, DS2) are run independently. Figure~\ref{fig: mean_grad_dissipation_production} shows the time series of the production and the dissipation from all the simulations. The production (and hence the dissipation at the statistically stationary state) from the non-Markovian field is closer to the DNS than the Markovian field.
Figure~\ref{fig: mean_grad_spectra_gradyc} shows the spatially averaged $|\nabla \phi|$ and the spectra of the scalar field, defined as ${C}(k) = |\hat{\phi}|^2/2$. As $\mathrm{Sc}$ decreases, the diffusive effects dominate the scalar mixing process. \textcolor{black}{Consequently, the scalar field length scale increases as $\sqrt{\mathrm{Sc}}$ (given $\mathrm{Re}$ remains constant), as highlighted in figures~\ref{fig: mean_grad_spectra_gradyc}b and d}. At the statistically stationary state, the scalar spectra for both NM16 and M16 cases approach the -17/3 scaling law derived by~\cite{batchelor1959small2}. This highlights the fact that the scalar spectrum exponent of -17/3 can be achieved even in a Markovian synthetic field as long as kinetic energy spectrum scales as $k^{-5/3}$.

\subsection{Decaying mixing}
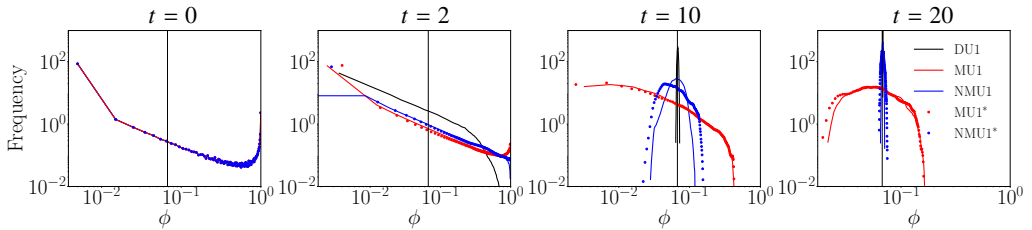
\begin{figure}
\centering
\includegraphics[width=\textwidth]{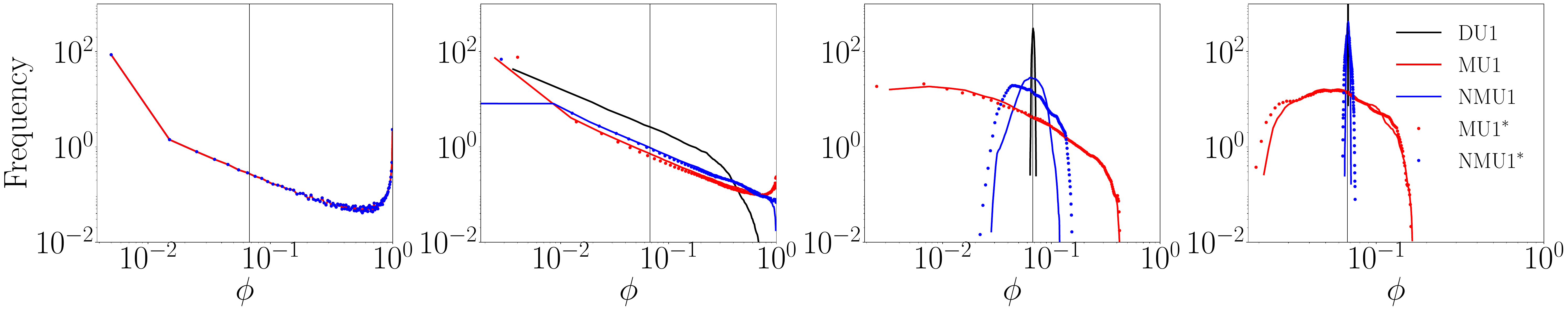}
 \put(-330,78){$t=0$}
 \put(-235,78){$t=2$}
 \put(-145,78){$t=10$}
 \put(-50,78){$t=20$}
\caption{PDF of concentration at different time instances for DNS, non-Markovian and Markovian velocity field for $\mathrm{Sc}=1$ cases.}
\label{fig: comb_pdf}
\end{figure}

To study mixing characteristics of a decaying case, we initialize the passive scalar concentration in a spherical blob defined as,
\begin{equation}
 2\phi(\bfx, 0) =\tanh \left(5\left(S - \pi/2\right)\right),
\end{equation}
where $S = \sqrt{(x-\pi)^2 + (y-\pi)^2 + (z-\pi)^2}$ and evolve using~\eqref{eq: scalar}. Figures~\ref{fig: mean_grad_spectra_gradyc}$c$, $d$ show the spatially averaged $|\nabla \phi|$ and scalar spectra $C(k)$ for all the decaying cases. The time evolution of $\avg{|\nabla \phi|}_x$ in the non-Markovian cases matches the evolution observed in the DNS. On the other hand, in all the Markovian cases, the scalar gradients are sustained over a very long period of time denoting very low stirring of scalar field by the velocity field. This is also evident from the spatial histogram of the scalar concentration shown in figure~\ref{fig: comb_pdf}, which highlight delayed homogenization due to Markovian fields, even though the scalar spectrum exponent approaches $-17/3$ for both the Markovian and the non-Markovian fields at large time. These results again show how non-Markovianity and the resulting spatiotemporal correlations of the synthetic velocity fields play a major role in the mixing process. 

\section{Conclusions}
\label{sec: Conclusions}
We have shown that non-Markovianity is an essential feature of turbulent flows, which can enhance turbulent mixing. Using an Ornstein-Uhlenbeck process with a different decorrelation time for each wavenumber, we demonstrated how a non-Markovian Gaussian velocity field can be used to improve the mixing properties of Markovian velocity models. Using the random-sweeping approximation, we use a decorrelation time inversely proportional to the wavenumber magnitude. We have shown that the time correlations of the synthetic velocity field thus obtained decay as $\tau^{-5}$ in a significant range of the time delay, thus confirming the non-Markovian nature of the velocity field. The Lagrangian velocity correlations, single particle correlations, and pair dispersions using the non-Markovian field are significantly closer to the DNS compared to the Markovian field. In the Eulerian description, scalars with all the Schmidt numbers exhibit better mixing properties (production in statistically stationary cases and mixing times in decaying blob mixing) with non-Markovian fields than the Markovian fields, though the exponent in scalar spectrum scales as -17/3 at low Schmidt numbers for both the fields.

Declaration of Interests. The authors report no conflict of interest.
\bibliographystyle{jfm}
\bibliography{jfm}

\end{document}